\documentclass[12pt,preprint]{aastex}

\slugcomment{}

\shorttitle{Ageism in Carina}
\shortauthors{Battaglia et al.}

\begin{document}


\title{The extensive age gradient of the Carina dwarf galaxy}


\author{G.Battaglia}
\affil{INAF - Osservatorio Astronomico di Bologna, via Ranzani 1, 40127, Bologna, Italy}
\email{gbattaglia@oabo.inaf.it}

\author{M.Irwin}
\affil{Institute of Astronomy, Madingley Road, Cambridge CB03 0HA, UK}

\author{E.Tolstoy, T. de Boer}
\affil{Kapteyn Astronomical Institute, University of Groningen, P.O.Box 800, 9700 AV Groningen, the Netherlands}

\and

\author{M.Mateo}
\affil{Department of Astronomy, University of Michigan, Ann Arbor, MI 48109-1090, USA}




\begin{abstract}
The evolution of small systems such as dwarf spheroidal galaxies (dSph) is likely to have been  
a balance between external environmental effects and internal processes within their own
relatively shallow potential wells.  Assessing how strong such environmental interactions may 
have been is therefore an important element in understanding the baryonic evolution of dSphs 
and their derived dark matter distribution. 

Here we present results from a wide-area CTIO/MOSAIC~II photometric survey of the Carina dSph, 
reaching down to about two magnitudes below the oldest main sequence turn-off (MSTO). This data-set 
enables us to trace the structure of Carina in detail out to very large distances from its center, 
and as a function of stellar age. 

We observe the presence of an extended structure made up primarily of 
ancient MSTO stars, at distances 
between 25\arcmin-60\arcmin\, from Carina's center, confirming results in the literature  
that Carina extends well beyond its nominal tidal radius. 

The large number statistics of our survey reveals features such as isophote twists and tails 
that had gone undetected in other previous shallower surveys. This is the first time that 
such unambiguous signs of tidal disruption have been found in a Milky Way ``classical'' dwarf 
other than Sagittarius.

We also demonstrate the presence of a negative age gradient in Carina directly from its MSTOs, 
and trace it out to very large distances from the galaxy center. The signs of interaction with 
the Milky Way make it unclear whether the age gradient was already in place before Carina 
underwent tidal disruption.

\end{abstract}





\section{Introduction}

The dwarf spheroidal galaxies (dSph) surrounding the Milky Way (MW) are the galaxies studied in better detail than any other 
\citep[see][for a recent review]{tht2009}, yet many fundamental 
questions about their nature remain unanswered. Are they the most dark matter dominated galaxies 
we know of, as their large apparent dynamical mass-to-light ratios would suggest, 
or is tidal disruption a contributing factor ?  
What drives the observed variety of star formation and chemical enrichment histories ?  Were dSphs born 
as we see them today or are they the descendants of larger and intrinsically different systems?

The evolution of small systems such as dwarf galaxies is likely to be particularly suspectible to the environment 
that they have experienced as well as being influenced by their own low potential well.  
However, the relative importance of both these factors has yet to be established.

Among the MW satellites, an example of a dSph whose evolution may have 
been strongly affected by the environment is given by Carina. This galaxy is the one ``classical'' dSph where most hints 
of tidal disruption have been found, such as the presence of probable members well beyond the nominal tidal radius, a
 rising velocity dispersion profile and a velocity shear reminiscent of tidally disrupted dwarf galaxies 
\citep[see][and references therein]{munoz2006}. However, no report 
of such kinematical features is given in \citet{koch2006} and \citet{walker2009}, so that the hypothesis of 
tidal disruption awaits confirmation. 

Carina is a particularly interesting object among the Local Group dwarf galaxies also because it 
is the only clear example of a galaxy with well-separated star formation episodes, 
visible as distinct main-sequence turn-offs (MSTO) in the Colour-Magnitude diagram 
\citep[e.g.][]{smecker-hane1996, monelli2003, bono2010}. 
The effects of its unique SFH can be traced in its chemical evolution as well \citep{koch2008, venn2012, lemasle2012}. 
It is unclear at present whether the influence of the MW may have induced 
the main SF episodes \citep[e.g. see simulations by][]{pasetto2011} as episodic SFHs 
can be qualitatively explained also by models of low mass dSphs treated in isolation \citep[e.g.][]{revaz2009}. 

Another feature that can provide information on the evolution of a small galaxy is the spatial distribution and kinematics 
of its stellar population mix. Negative age and metallicity gradients are found in several of the Local Group dSphs, both 
around the MW \citep{harbeck2001, tolstoy2004, battaglia2006, battaglia2011, deboer2012a, deboer2012b} and in isolation 
\citep[e.g.][]{bernard2008}; the latter finding suggests that internal effects are the main drivers 
in setting up the observed gradients. In Carina, indications of an age gradient have been found from the spatial distribution 
of red clump (RC) stars as compared to horizontal branch (HB) stars \citep{harbeck2001}. 
Although this is similar to the behaviour of other MW dSphs with multiple stellar components, 
for Carina there have been no findings of 
different kinematics among the stars belonging to different age/metallicity ranges \citep[e.g.][]{walker2011}, 
unlike what observed in Fornax, Sculptor and Sextans \citep{tolstoy2004, battaglia2006, battaglia2011, walker2011}. 
Interestingly, it has been shown that strong interaction with the MW is 
a viable mechanism for erasing differences in the kinematics of multiple stellar components \citep{sales2010}.

Several of the observed properties of Carina could therefore be explained in a framework of strong 
interaction with the MW. It is then particularly interesting to fully characterize the stellar population of this object 
and to put on a secure ground the possible existence of tidal features. 

Here we present results from a deep, wide-area photometric survey of Carina, which 
yields major improvements with respect to previous studies of the morphology and spatial distribution of Carina's stellar 
population mix. 

\section{Observations}
Deep optical photometry covering the Carina dSph was obtained using the CTIO 4-m MOSAIC~II camera over 6 nights in December 2007, 
as part of observing proposal 2007B-0232~(PI: M. Mateo). Observations in the B and V bands were obtained in 8 pointings around the 
centre of Carina, covering a total area of $\approx$2 deg$^{2}$. The observing strategy was to obtain multiple long~($\approx$600s) 
exposures for each pointing, which have been stacked together to obtain deep photometry. In addition, short~(30s) exposures 
were also obtained, to sample bright stars that are saturated in the deep images. 
To ensure accurate photometric calibration of the data set, observations of Landolt standard fields~\citep{landolt1992} 
were made, covering a range of airmass and colour. 

First stage image processing steps included standard bias and overscan-correction, 
trimming to the reliable active detector area, and flatfielding via creation of master flats from a 
dithered set of B-band and V-band twilight
sky exposures. The flatfielding step also corrects for internal gain
variations between the detectors.

Prior to deep stacking of the individual exposures, detector-level catalogues were generated for each
science image to refine the astrometric calibration and also to assess the
data quality. For astrometric calibration, a Zenithal polynomial projection \citep{greisen2002} 
was used to define the World Coordinate System. 
A third-order polynomial includes all the significant telescope
radial field distortions leaving just a six-parameter linear model per
detector to completely define the astrometric transformations. The
TwoMicronAll Sky Survey (2MASS) point-source catalogue \citep{cutri2003} was
used for the astrometric reference system.

The common background regions in
the overlap area from each image in the stack were used to compensate for
sky variations during the exposure sequence and the final stack included
seeing weighting, confidence (i.e. variance) map weighting and clipping
of cosmic rays.

Catalogues for each deep stacked image were then generated using standard
aperture photometry techniques \citep[e.g.][]{irwin2004} and cross-matching
with published photometry of Carina \citep{stetson2000} was used to directly
photometrically calibrate the central fields.  Outlying fields were
calibrated with respect to the central region using the overlaps
between adjacent pointings.




\section{Results} \label{sec:res}

\subsection{Age Gradient} \label{sec:cmds}

While the exact age of Carina's SF episodes varies 
among the various works in the literature, there is consensus 
about Carina experiencing at least three main SF episodes: 
an ``ancient'' one, around 
11-13 Gyr ago; a main episode at intermediate ages \citep[around 2.5-4 Gyr ago][but 
$\sim$7 Gyr ago according to Hurley-Keller et al. 1998]{bono2010}, after a long quiescent period; 
a relatively recent 
episode, $\ga$1 Gyr ago \citep[e.g.][]{monelli2003}.

The way these ``ancient'', ``intermediate-age'' and ``young'' stellar components 
are spatially distributed has not been studied in detail yet. 
Notably, notwithstanding its complex SFH, Carina has a very narrow red giant branch (RGB).
This makes  
photometry of RGB stars unsuitable for exploring the existence of age gradients in this galaxy, 
even though such spatially extended data-sets exist \citep[$\sim$10deg$^2$][]{munoz2006}. 
Indications for a negative age gradient in Carina were found by 
\citet{harbeck2001} over a relatively small radial range (30\arcmin), using photometry down to $\sim$1mag below the 
HB. 
The shallow photometry allowed only the use of indirect age 
indicators for tracing the gradient, 
i.e. the RC for the young/intermediate-age stars and the HB for the ancient stars. 

\begin{figure}
\epsscale{.90}
\plotone{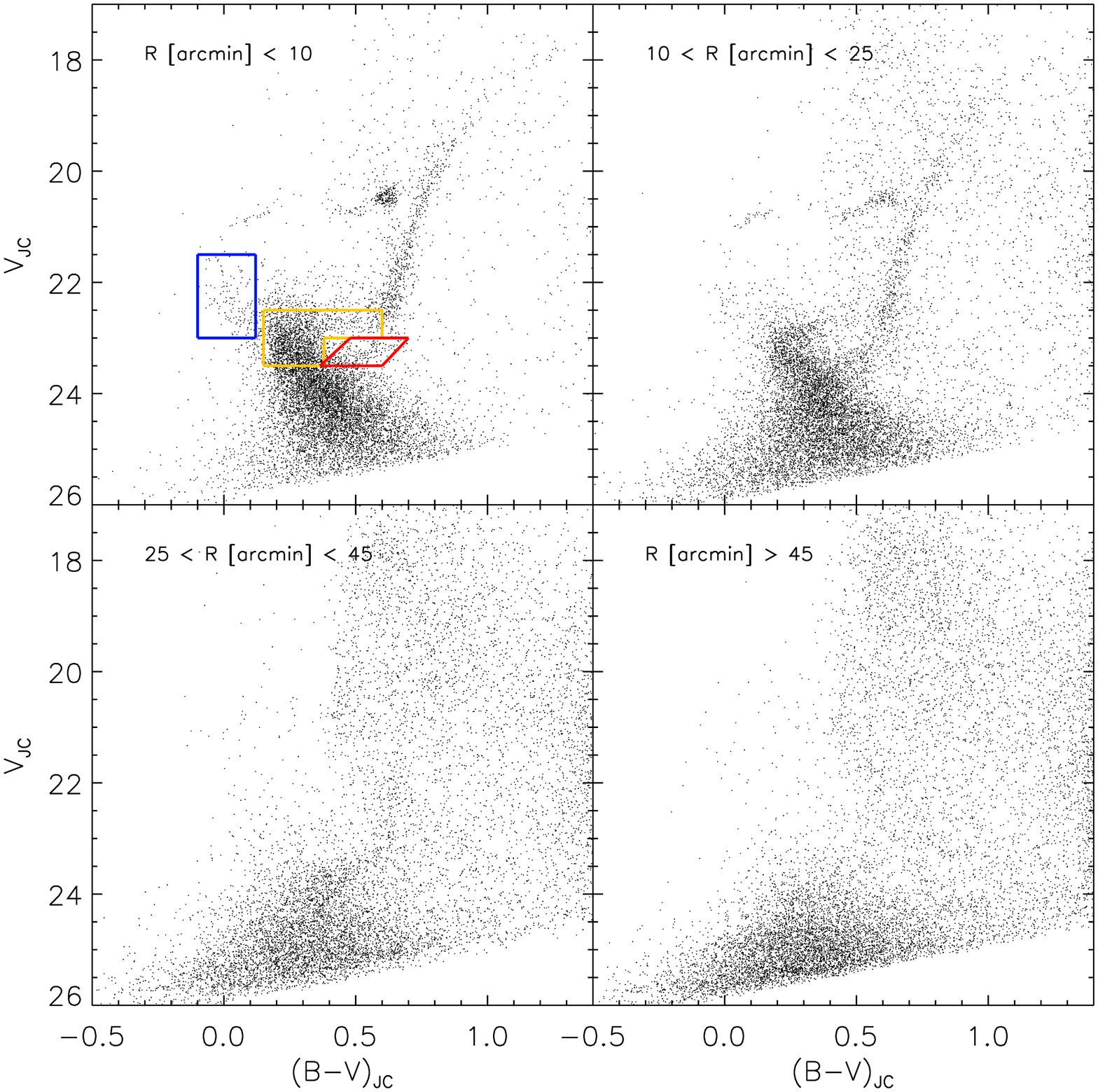}
\caption{
Colour-magnitude diagram of stellar objects along the line-of-sight to the Carina dSph, over different ranges of 
projected elliptical radius (see labels). The elliptical annuli have P.A. and $e$ values from IH95. All the 
panels contain the same number of stars (15000) to enable a comparison of the 
relative strength of the features in the CMD. The boxes 
in the top-left panel show the cuts in magnitude and color applied to select young (blue), intermediate-age (yellow) and old (red) stars 
for Figure~\ref{fig:fov}.} \label{fig:cmd}
\end{figure}

Figure~\ref{fig:cmd} shows the Colour-Magnitude Diagram (CMD) resulting from our photometric survey, 
split over a selection of elliptical radii $R$. This spatially extended 
photometric data-set reaches down to $\sim$2 magnitudes below the oldest 
MSTO, allowing us to trace possible age gradients {\it directly} from the 
presence or absence of the MSTOs belonging to the various SF episodes {\it and} over a large area.  

In the innermost region (R $<$ 10\arcmin) all the main features of the Carina stellar population 
are present: the thin RGB, the RC, the red and 
blue HB, the ancient MSTO
 and sub-giant (SGB), the intermediate-age MSTO and SGB, 
and younger MS stars at \bv$\sim$0 and 21$\la$ {\it V} $\la$ 23 \citep[for a detailed 
discussion on the various features of the CMD see e.g.][]{hurley1998}.  
At larger radii (10\arcmin\,$<${\it R}$<$ 25\arcmin) the young stars are barely visible, while the 
features containing intermediate-age and ancient stars are still well 
populated. In the outer parts ({\it R}$>$ 25\arcmin), the MSTO of the intermediate-age 
population disappears, while the ancient stars are still visible, 
as witnessed by the presence of the old MSTO and SGB. 

The above direct detection of an age gradient in Carina, as traced from MSTO stars,  
confirms the early findings of \citet{harbeck2001} and extends them over a much larger radial range. 

We note that the presence of a sizeable population of 
ancient MSTO stars in the last two panels of Figure~\ref{fig:cmd} ({\it R}$>$ 25\arcmin\,) shows that 
Carina extends well beyond its nominal King tidal radius \citep[$r_t = $28.8\arcmin\,,][hereafter IH95]{irwin1995}. 
This puts on a secure ground the conclusions by \citet{munoz2006} based on 
the spatial distribution and kinematics of a smaller sample RGB stars.

\begin{figure}
\epsscale{.50}
\plotone{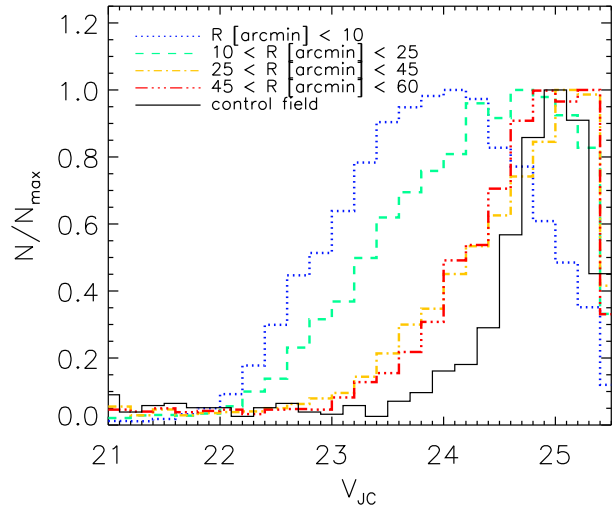}
\caption{
Luminosity function of Carina's stars, normalized to the peak value and plotted in bins of elliptical 
radii (see legend). The stars are  
selected to have 21 $<$V$<$25.5, B-V$<$0.6. 
The control field corresponds to the region with rectangular coordinates 
60\arcmin $< \xi <$ 80\arcmin and 
20\arcmin $< \eta <$ 40\arcmin (see Fig.~\ref{fig:fov}).} \label{fig:lf}
\end{figure}

Both the presence of the age 
gradient and the large extent of Carina are also illustrated by Figure~\ref{fig:lf}, where we compare to a control field 
the luminosity function (LF) in bins of $R$ of stars in the region of the MS, MSTO and SGB. 
Owing to the large extent of Carina it is difficult to accurately 
identify a region containing only contaminants (i.e. MW stars and unresolved galaxies) to use as a control field. 
We chose to use a sub-region of the outermost pointing on 
the North-East (see Figure~\ref{fig:fov}) because it has the lowest density of objects among the various pointings; the significantly 
different shape of the control field LF from the two outer elliptical zones confirms that it mostly contains 
contaminants and demonstrates the presence of appreciable numbers of Carina members out to 1 degree radius. 
Furthermore, the similarity of the LFs of the two outer
zones and the absence of obvious intermediate-age and young stars beyond
$\sim$25\arcmin\, strongly suggest this extended structure is made up primarily
of ancient stars. 

\subsection{Extended Spatial Structure} \label{sec:tid}

\begin{figure}
\centering
\begin{tabular}{cc}
\includegraphics[width=0.5\linewidth]{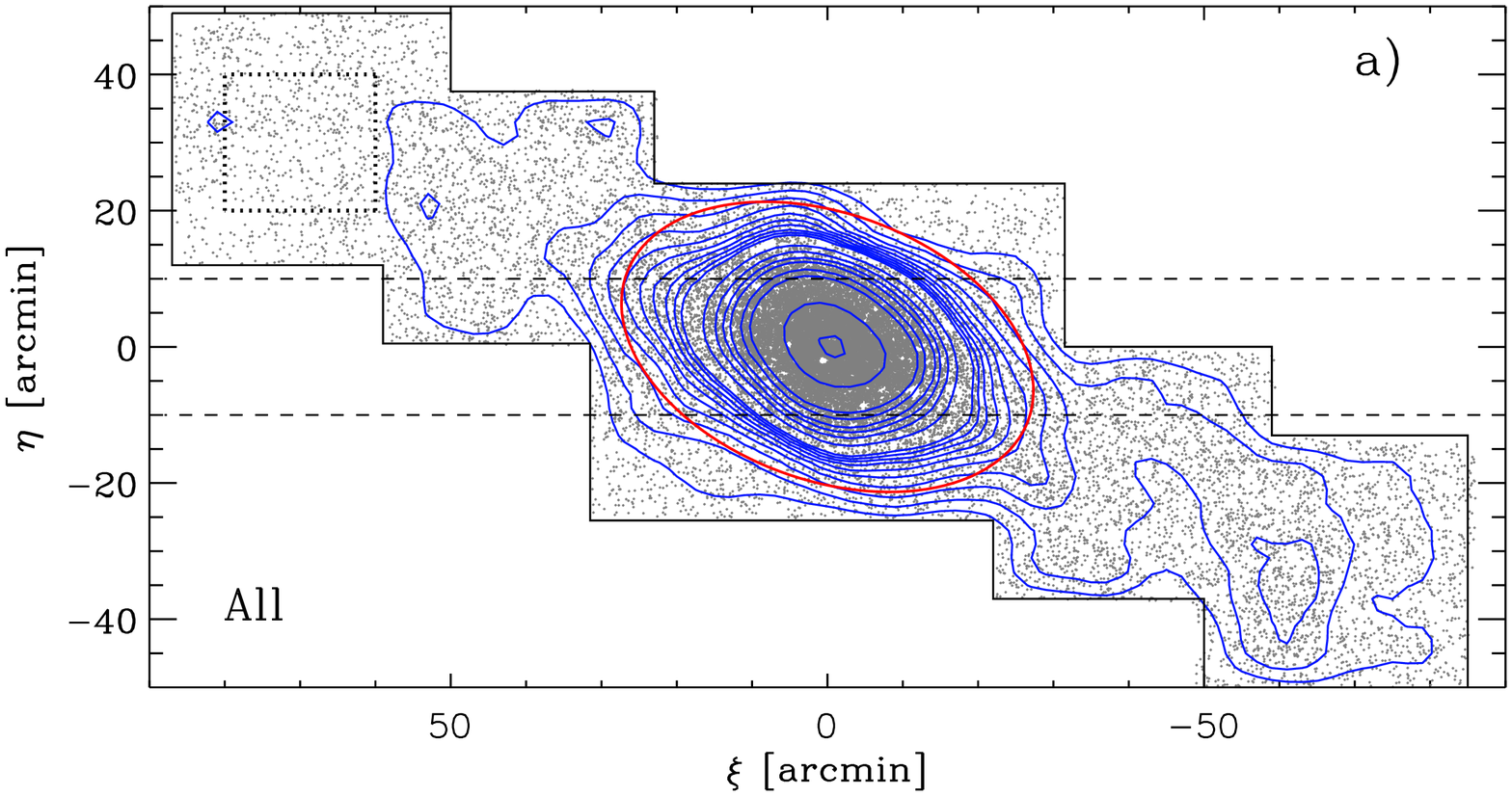} & 
\includegraphics[width=0.5\linewidth]{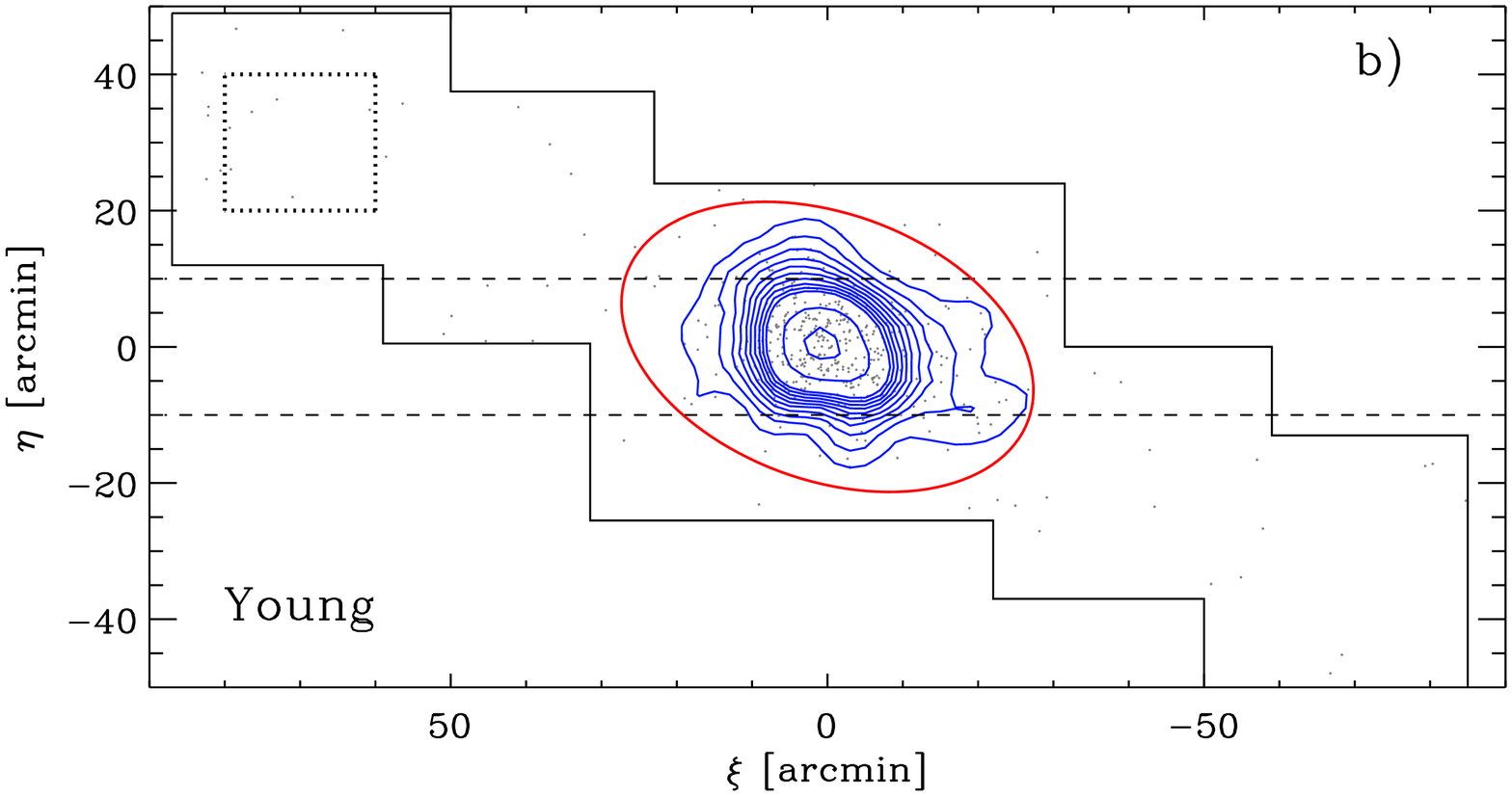} \\
\includegraphics[width=0.5\linewidth]{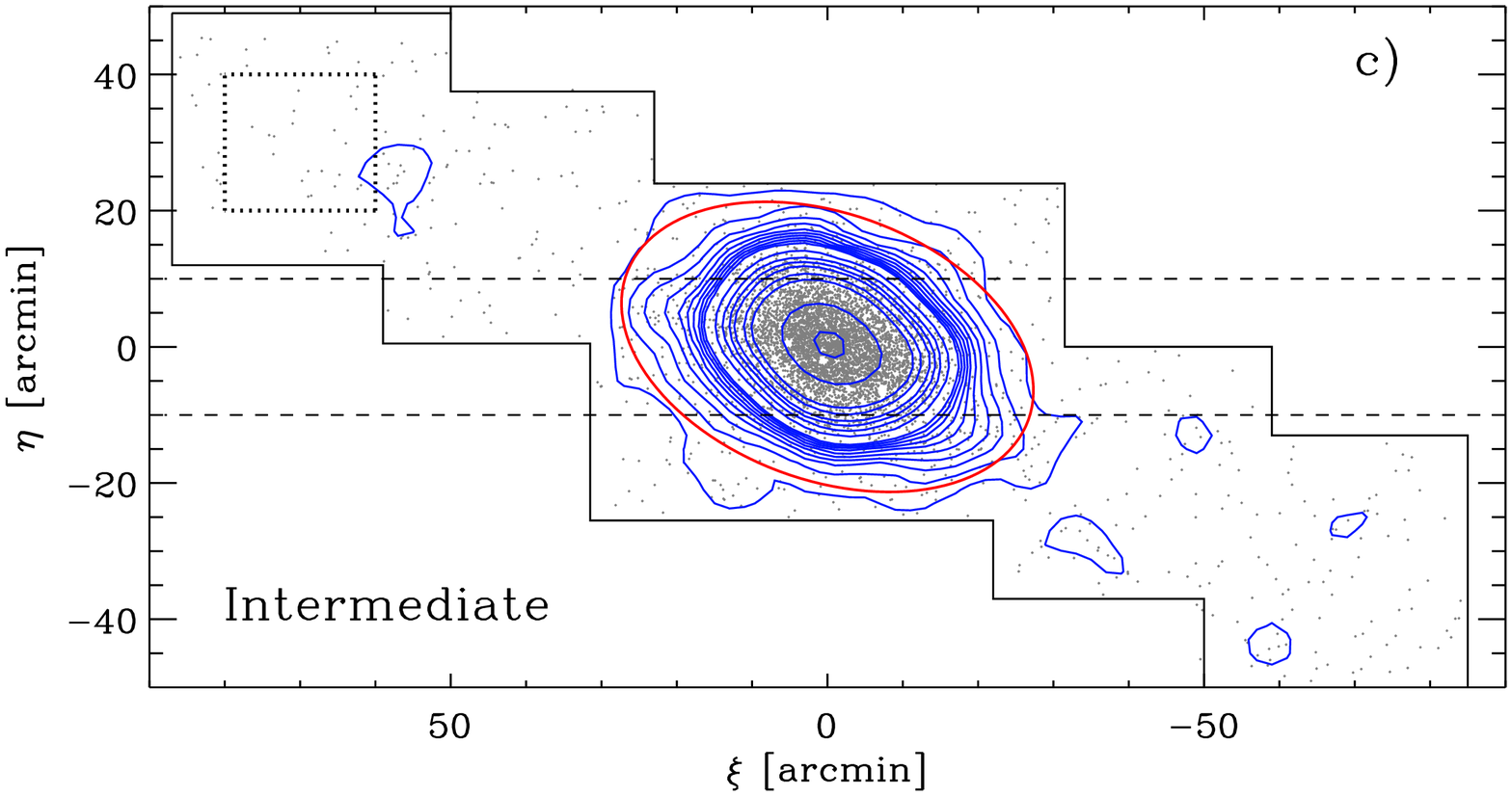} & 
\includegraphics[width=0.5\linewidth]{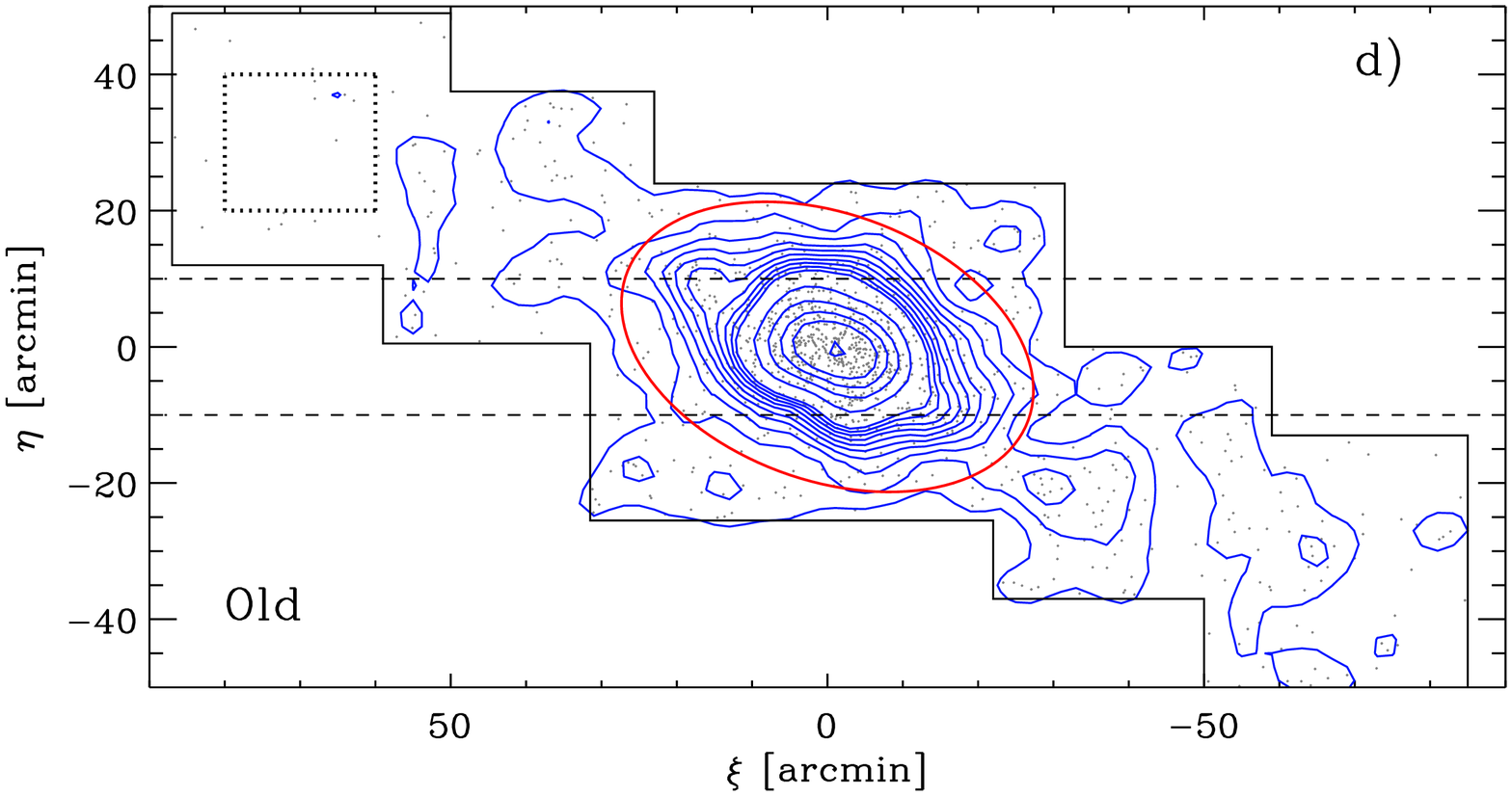} \\
\end{tabular}
\caption{Spatial distribution of stars in Carina for the overall stellar population, 
young stars, intermediate age population, old stars (see labels and Figure~\ref{fig:cmd}). The overlaid iso-density contours 
were derived from 
the contaminant-density subtracted Hess diagram obtained on a 2\arcmin$\times$2\arcmin pixel grid, and show 
levels of 1,2,3,4,5,6,7,8,9,10$\sigma$,15,20$\sigma$... (where $\sigma$ is the noise, see text for details). The values of 
$\sigma$ are 2, 0.51, 0.24, 0.24 stars/pixel for panels a,b,c,d, respectively, while the values of the subtracted $\Sigma_{\rm cont}$ are 
3.08, 0.007, 0.01, 0.01 stars/pixel; the region used for determining $\sigma$ and $\Sigma_{\rm cont}$ is enclosed by the dotted lines. 
The ellipse indicates the nominal tidal radius (parameters from IH95; center coordinates from Mateo 1998). The horizontal dashed lines 
are meant 
to guide the eye when comparing the direction of the isophote elongation for the various populations. 
} \label{fig:fov}
\end{figure}

Figure~\ref{fig:fov}a illustrates the spatial distribution of the overall Carina stellar population 
after applying magnitude cuts of B$<$25.1, V$<$25.5 - corresponding to $\sim$50\% completeness levels - to counter
the variable depth of the pointings due to differences in total exposure time. 
To minimize the presence of contaminants, we isolate the main features of Carina's overall stellar population on the 
CMD via selection in magnitude and color. Figure~\ref{fig:fov}b,c,d show how the spatial distribution varies with age. 

The isodensity contours in each panel were derived from maps of stellar counts (Hess diagrams) 
on a 2\arcmin\, $\times$ 2\arcmin\, pixel grid, from which we subtracted the density of contaminants, $\Sigma_{\rm cont}$. 
This was calculated as the median of the pixel values in the control 
field (shown in Fig.~\ref{fig:fov} and discussed in \ref{sec:cmds}). The iso-density contours are then plotted in units of noise 
($\sigma$), which we determine as the r.m.s. of the 
pixel values in the contaminant-substracted control field. 
Should the derived $\Sigma_{\rm cont}$ be overestimated because of a leakage 
of Carina's stars, the main effect will cause Carina to appear 
smaller than it is and to underestimate the significance of the contours; hence the extent of Carina 
found here shold be considered as a ``lower limit'' and the significance of the outer contours as conservative.

Figure~\ref{fig:fov}a shows that the main body of Carina 
appears of regular morpology, although with slightly boxy iso-density contours, 
out to {\it R}$\sim$18-20\arcmin\, $\sim$ 0.6-0.7 r$_{\rm t}$. 
Beyond this radius, the contours become elongated and distort, 
with very extended tails emerging from both sides of the main body and a
well-defined over-density of stars 
at ($\xi$,$\eta$)$\sim$($-$60\arcmin,$-$30\arcmin). The much larger number statistics of this study with respect 
to studies in the literature makes it possible to observe isophote elongation and twists; these had not be 
detected in previous, shallower photometric surveys of Carina, even though optimized to minimize 
the contamination from MW stars \citep[e.g.][]{munoz2006}.

Fitting of ellipses as function of $R$ to the map of stellar counts confirms the change in shape and 
orientation of the contours, yielding a P.A. and ellipticity $e$ that are constant around a P.A. 
$=$ 53.5\degr\,$\pm$1.0\degr\, and $e=$ 0.29$\pm$0.01 
at $R <$ 18\arcmin\,, and then increase, yielding a P.A.$=$ 62.5\degr\,$\pm$0.5\degr\, and $e=$ 0.35$\pm$0.02 at 
$R >$ 18\arcmin\,; the latter values tend 
to the literature values of P.A.$=$65\degr\,$\pm$5\degr\,, $e=$0.33$\pm$0.05 \citep{irwin1995}.

\begin{figure}
\centering
\begin{tabular}{cc}
\includegraphics[width=0.5\linewidth]{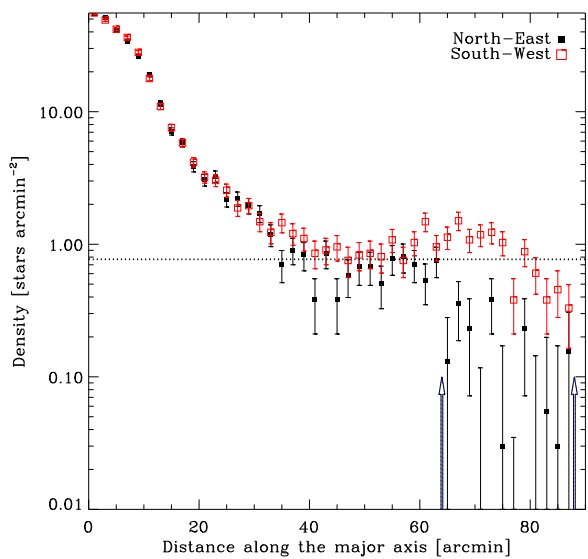} \\
\end{tabular}
\caption{Surface number count profile of Carina, derived from the stars in Figure~\ref{fig:fov}a) that lie on a 20\arcmin\, wide 
slice running along a P.A.= 65\degr. Filled and open squares refer to the North-East and South-West part of the galaxy, respectively. 
The contaminant density subtracted from the profile is indicated by the dotted line; the arrows show the radial range occupied by the 
region for calculating $\Sigma_{\rm cont}$. We note the change of 
slope at R$\sim$20\arcmin\, present in both halves of the galaxy, and the overdensity at R$\sim$70\arcmin\, in the S-W half, corresponding to the 
overdensity visible in Figure~\ref{fig:fov}a) at ($\xi$,$\eta$)$\sim$($-$60\arcmin,$-$30\arcmin). 
} \label{fig:dens}
\end{figure}

We note also the presence of a break in the surface number count profile 
of Carina's stars lying along the projected major axis of the system 
(Figure~\ref{fig:dens}). The break occurs in both N-E and S-W halves of the 
galaxy at around 20\arcmin from the center, where the iso-density contours 
start becoming elongated, and at a similar radial distance to that found
by \citet[e.g.]{majewski2000, walcher2003, munoz2006} in shallower samples.

The youngest Carina stars do not extend as far out as the intermediate-age 
and ancient ones, but signs of isophote elongation are arguably also visible 
in the spatial distribution of these stars (Figure~\ref{fig:fov}~b). This would 
set the last pericentric passage less than 1-2 Gyr ago. The intermediate-age 
and ancient population both show distorted outer isophotes 
(Figure~\ref{fig:fov}~c,d). 
Notably, as visible in the figure and confirmed by ellipse fitting of the Hess diagrams, 
the shape parameters of the contours are 
subtly different for these two stellar populations of different ages and may
impact on models trying to reconstruct the orbital history of Carina. 

\section{Discussion and conclusions}

We have used CTIO/MOSAIC~II photometry to study the morphology and spatial distribution of Carina's stellar 
population mix.  

The combination of wide-area and photometric depth, reaching down to $\sim$2~mag below the oldest MSTO, 
allows for 
a direct detection of an extensive age gradient in Carina, traced from the presence/absence of the MSTOs 
belonging to the various SF episodes. 

We find a clear presence of ancient MSTO stars at $R>$ 25\arcmin\,, which demonstrates that 
Carina extends well beyond its literature nominal tidal radius, confirming results from  
previous works. 
The increased number statistics from this study reveals the existence of 
faint features such as isophote elongation and twists, that could not be detected in previous 
shallower photometric surveys. 
Such features are even visible when separating the stars in age. We confirm the presence
of a clear break in the surface brightness profile at radii $>$20 \arcmin\, and observe 
an over-density of stars 
in the S-W half of Carina, at $\sim$70\arcmin\, from the center. 

Features such as breaks in the surface brightness profile, together with variations 
in the elongation and direction of 
the isophotes, have been detected in other objects that are good candidates for
ongoing tidal distruption such as NGC205 and M32 \citep{choi2002}, and are 
predicted by N-body simulations of tidally disrupted dwarf galaxies 
\citep[e.g.][]{choi2002, munoz2008, penarrubia2009}. These morphological 
traits strongly suggests that Carina is being tidally disrupted. 

At a distance of $\sim$100kpc the low Galactocentric line-of-sight velocity of 
Carina ($\sim$7 km/s) places it either close to apocenter or pericenter. Given
the signs of tidal disturbance and the current best estimate of the proper 
motion \citep{piatek2003}, Carina is most probably close to apocenter with 
an orbital period of between 1-2 Gyr and hence the last pericentric passage
likely occurred  $<$1 Gyr ago. The presence of signs of tidal disruption at apocenter is 
consistent with the timescales from the simulations of \citet{penarrubia2009}, where 
tidal tails can remain clearly visible at projected radii larger than the break 
radius until up to $\sim$15 dynamical crossing times, i.e. between 1-2 Gyr 
in this case.

The existence of signs of tidal disruption complicates the interpretation of the observed age gradient, as 
it is not obvious whether it was (in part) already in 
place before Carina was tidally disturbed.
Related questions are whether 
Carina showed distinct kinematic signatures for the different populations
prior to tidal interactions and how these may have been affected by close
pericentric passages.

This also impacts on the derived dark matter content and distribution.
Treated as an isolated equilibrium system, Carina has a large dynamical 
mass-to-light ratio 
\citep[e.g. M/L$_{V} \sim$100 (M/L)$_{V, \odot}$,][]{walker2007}, 
as do most other ``classical'' dSphs. An attempt to model Carina as a tidally 
perturbed system was made by \citet{munoz2008}, who found that the surface 
brightness and internal kinematic properties of Carina are 
compatible with a mass-follows-light model still containing large amounts of dark matter 
(M/L$\sim$40). The 2D distribution 
resulting from such model is remarkably regular, in contrast with what is
observed here over the same spatial region.  However, it should be considered that 
the best model of 
\citet{munoz2006} was aiming at reproducing the spatial distribution from the data available at the time, 
too shallow to reveal the signs of tidal distorsion seen in our survey.

Knowledge of the distorted structure of the galaxy from this deep survey, 
combined with the extensive spectroscopic data-sets present in the literature 
\citep[e.g.][]{koch2006, munoz2006, walker2009}, should provide important 
constraints for assessing the impact of tidal disruption on the derived 
mass-to-light ratio - and dark matter content.  

Tidal disruption of dwarf galaxies has been detected in some M~31 satellites 
\citep[e.g.][]{choi2002, geha2006, mcconnachie2006}. However, around the MW, 
apart from the clear case of Sagittarius, the quest has proven difficult  
due to the combination of the large extent on the sky of MW dSphs and their 
low surface brightness. The question therefore remains if tidal 
disruption is common among the MW dSphs and 
ultra-faint dwarfs, or whether Carina belongs to a  ``minority'' of tidally 
disrupted objects such as the Sagittarius dwarf \citep{ibata1994}.
Hints of tidal disruption have been found in some of these other systems, e.g. 
Ursa~Minor \citep[e.g.][]{martinez-delgado2001, palma2003}, Leo~I \citep{mateo2008}, 
Ursa~Major~II \citep{munoz2010}. In contrast, Draco, the only other dSph 
surveyed with photometric data as deep and 
wide-area as those used here, shows a ``flawless'', regular 
morphology \citep{segall2007} and an enourmous M/L $\sim$300 \citep{kleyna2002}.


\acknowledgments
The research leading to these results has received funding from the European 
Union Seventh Framework Program (FP7/2007-2013) under grant agreement number PIEF-GA-2010-274151. 
The authors acknowledge the International Space Science Institute at Bern for their funding of the team ``Defining 
the full life-cycle of dwarf galaxy evolution: the Local Universe as a template''. 



{\it Facilities:} \facility{Blanco (Mosaic~II)}




\clearpage


\clearpage

\end{document}